\documentclass{evn2004}
\usepackage{txfonts}
\begin{document}
   \title{VLBI polarimetric observations of 3C147}

   \author{A. Rossetti\inst{1} \and
           F. Mantovani \inst{1} \and
           D. Dallacasa \inst{1,2}
            }

\institute{
Istituto di Radioastronomia del CNR, via Gobetti 101, I-40129, Bologna, Italy \and 
Dipartimento di Astronomia, Universit\`a degli Studi, via Ranzani 1, I--40127 Bologna, Italy }

   \abstract{We present multi-frequency VLBI observations of the Compact Steep-spectrum Quasar 
    3C147 (0538+498) made with the VLBA at the four frequencies in the available 5~GHz band and 
    at 8.4~GHz (still under analysis), from which we derived millarcsecond-resolution images 
    of the total intensity,
    polarization, and rotation measure distributions. The source shows a core-jet structure, with
    a compact feature and a jet, extending about 200 mas to the South-West. We detect polarized 
    emission in two bright features in the inner jet; the rotation measure of this features 
    ($\sim$ -1630~rad~m$^{-2}$, $\sim$ -540~rad~m$^{-2}$).}

   \maketitle
%

\section{Introduction}
Compact Steep-spectrum Sources (CSSs) seem to be a scaled-down version of large-sized
double radio sources. They have linear sizes $<20~h^{-1}~kpc$ ($q_{0}=0.5$ and 
$H_{0}=100~h$~km~s$^{-1}$~Mpc$^{-1}$) and steep high-frequency radio spectra.
It is now generally believed that most of them are {\it young} radio sources, with
life-time $<10^{3-5}$~year, whose radio lobes have not had time to grow to kilo-parsec
scales (Fanti et al. \cite{Fanti95}).
 
Among them there is a subclass of quasars characterized by very luminous jets which often show
polarized emission and high integrated Faraday rotation.

Owing to their small sizes these objects are fully immersed in the host galaxy ISM and their 
distorted, complex morphologies (like in 3C119, 3C343) show evidence of strong 
interaction of the jets with the dense ambient medium.

Here we report VLBA polarimetric observations at 5~GHz of the CSS quasar 3C147.
These observations are part of a program aimed at studying the polarization characteristics
of the core regions, jets and lobes and possibly the jet-cloud interactions in CCSs.  

\section{General information}

3C147 (0538+498) is a prominent CSS source at $z=0.545$ with a total angular extent of 0.6
arcsecond, corresponding to a projected linear size of $\sim 2.2~h^{-1}~kpc$.

On the sub-arcsecond scale it displays a very asymmetric double-lobed radio structure
with a prominent central component (Akujor \& Garrington \cite{Akujor95}; Junor et al. 
\cite{Junor99}) which is resolved out from VLBI observations into a compact core with a jet
emerging from it toward the South-West (Alef et al. \cite{Alef90}).

This object shows a very large integrated rotation measure, RM$\sim-1500$~rad~m$^{-2}$, (Inoue et 
al. \cite{Inoue95}), suggesting that the radio source may be surrounded by a dense ionized medium.

\section{Observational results}

\begin{figure}
   \centering
   \includegraphics{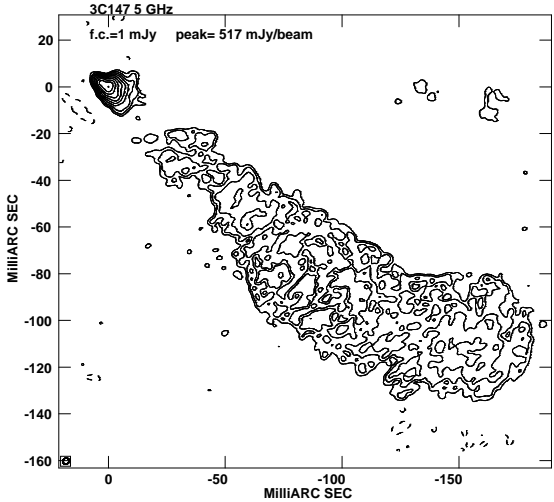}
   \vspace{6.5cm}
   \caption{\footnotesize 5~GHz image of 3C147 with a restoring beam of 2.8 x 2.5 mas
            at -40$\degr$ and 4 IFs.
            Contour levs increase by a factor of 2.}
            \label{fig:total}
           
    \end{figure}

\begin{figure}
   \centering
   \includegraphics{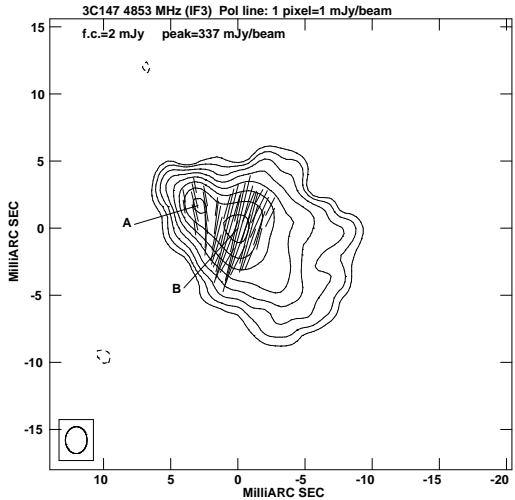}
   \vspace{6.5cm}
   \caption{\footnotesize 4.85~GHz image of the central compact component of 3C147 with
            polarization angle sticks for the IF 3 superimposed.
            Contour levs increase by a factor of 2.}
            \label{fig:vettori}
          
    \end{figure}

We observed 3C147 for about 12 hours at 5~GHz and 8.4~GHz with the VLBA in May 2001. 
The data were correlated with the NRAO VLBA processor at Socorro and calibrated, imaged and analysed
 using the AIPS package.
Images obtained from our data are shown in Fig.~\ref{fig:total}--~\ref{fig:vettori}.
The rms noise level is about 0.2~mJy, but we use a first contour level greater than $3\sigma$
because of the presence of minor but significant 
residual calibration errors, probably due to the difficulty in imaging complex structures having both 
compact features and very extended emission.

On the mas-scale 3C147 shows a one-sided
core-jet structure. The observed structure can be divided into two parts: the complex 
compact region (which probably contains the core and the head of the jet) and the jet,
extending to the South-West, which is followed out to a distance of about 200~mas from
the compact component. It widens soon without loosing its collimation, shows several gentle
 wiggles and turns toward the North near the end of its length.
No signs of counter-jet are seen.
Fig.~\ref{fig:total} shows two compact components embedded in a more diffuse emission:
one at the position of the peak (B) and one to the North-East of the peak (A) as in
Nan et al. (\cite{Nan00}). We find a total flux density of 4.70 Jy.
Total and polarized flux densities integrated on the same regions
for the compact components are
given in Table~\ref{tab:fluxes}. Polarized emission has been detected from both components.

\begin{table}[htbp]
\centering
 \caption{Source parameters.}
 \label{tab:fluxes}
 \begin{tabular}{cccccc}
 \hline
 \hline
Component  &  IF  &   S   &   $p$  &  $\chi$   & \%pol    \\
           &  MHz &  mJy  &   mJy  &  $\degr$  &         \\
\hline
\hline
  A        & 4619 &  345  &   3.3  &   8.4     & $\sim1.0$\\      
           & 4657 &  332  &   3.0  &  -1.0     & $\sim0.9$\\
           & 4854 &  339  &   3.1  &  12.3     & $\sim0.9$\\
           & 5090 &  330  &   2.9  &  27.2     & $\sim0.9$\\ 
           &      &       &        &           &          \\
  B        & 4619 &  707  &   9.6  &  -46.5    & $\sim1.4$\\
           & 4657 &  688  &  11.6  &  -44.5    & $\sim1.7$\\
           & 4854 &  693  &  13.6  &  -13.3    & $\sim2.0$\\
           & 5090 &  671  &  11.6  &  +21.8    & $\sim1.7$\\
\hline
\hline
\end{tabular}
\end{table}

\begin{figure}
   \centering
   \includegraphics{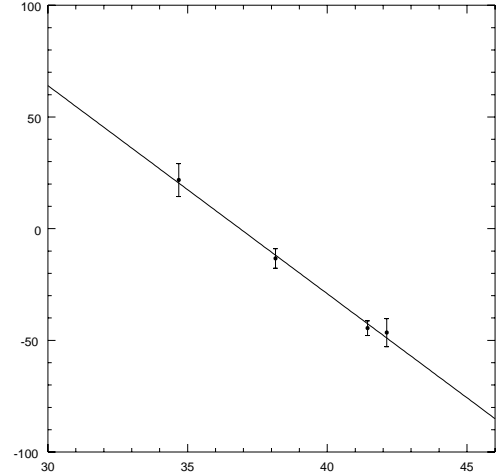}
   \vspace{6.5cm}  
   \caption{\footnotesize Plot of the observed $\chi$ values for the polarization of component B
            as a function of $\lambda ^{2}$ for the four wavelengths.}
            \label{fig:plot}
           
    \end{figure}

\begin{figure}
    \centering
    \includegraphics{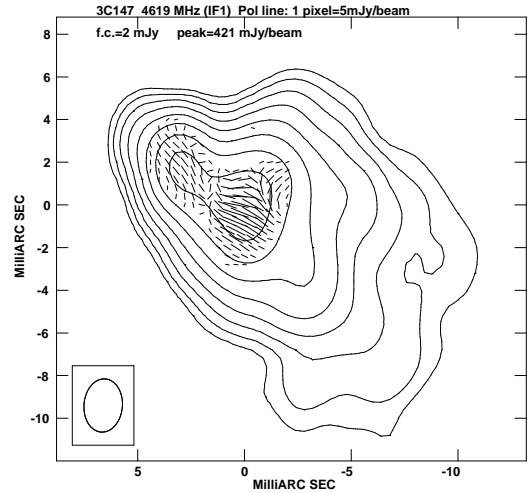}
    \vspace{6.5cm}  
    \caption{\footnotesize 4.6~GHz image of the central compact component of 3C147 with
             the intrinsic orientation of the magnetic field vectors superimposed.
             Contour levs increase by a factor of 2.}
             \label{fig:vettoriB}
           
     \end{figure}

In Nan et al. (\cite{Nan00}) there are indications that component A is not polarized and, based on its
weak polarization, they identify it as the core. The spectral index we derive for component A
 is $\alpha_{5}^{8.4}\sim 0.3$.
We suggest that the polarized emission we detect at 5~GHz for component A comes from 
the head of the jet and that we have not enough resolution to resolve the real core.
In Table ~\ref{tab:fluxes} we give the polarization angle values at each of the four 
frequencies. We used that values at the peak of the polarized flux density of the four $p$
maps to estimate the RM for components A and B. A best squares fit to this data gives 
an integrated RM of $-1630\pm 110$
~rad~m$^{-2}$ for component B (see Fig.~\ref{fig:plot}) and of $-540\pm110$~rad~m$^{-2}$ for 
A component, where uncertainties due to both the calibration and the image noise levels have been considered.
No redshift corrections have been applied to the wavelength, so the RMs in the
 rest of frame of 3C147 results of about $-3900\pm260$ ~rad~m$^{-2}$ for the B component and $-1290\pm260$
~rad~m$^{-2}$ for the A component.
In order to establish the orientation of the magnetic field we have extrapolated the polarization angle to 
$\lambda=0$ and have found $\chi\sim-16\degr\pm23\degr$. As a result, the magnetic field vectors should be oriented
along the direction of the jet (see Fig.~\ref{fig:vettoriB}),
but a more accurate study of the magnetic field orientation could be done only when 
the 8.4~GHz information will be available.    

\begin{acknowledgements}
The VLBA is operated by the U.S. National Radio Astronomy Observatory which is a facility of the National
Science Foundation operated under a cooperative agreement by Associated Universities, Inc.
\end{acknowledgements}

\end{document}